\documentclass[aps,twocolumn,prl,showpacs,floatfix,superscriptaddress]{revtex4}
\usepackage{graphicx}
\usepackage{bm}
\usepackage{subfigure}
\usepackage{amssymb}
\usepackage{amsmath}

\begin{document}

\title{Anomalous Excitation Spectra of Frustrated Quantum Antiferromagnets}

\author{Weihong Zheng}
\affiliation{School of Physics, The University of New South Wales,
Sydney, NSW 2052, Australia}
\author{John O. Fj{\ae}restad}
\affiliation{Department of Physics, The University of Queensland,
Brisbane, QLD 4072, Australia}
\author{Rajiv~R.~P.~Singh}
\affiliation{Department of Physics, University of California, Davis,
CA 95616, USA}
\author{Ross H. McKenzie}
\affiliation{Department of Physics, The University of Queensland,
Brisbane, QLD 4072, Australia}
\author{Radu Coldea}
\affiliation{Department of Physics, University of Oxford, Oxford OX1
3PU, United Kingdom}

\date{\today}


\begin{abstract}
We use series expansions to study the excitation spectra of
spin-$1/2$ antiferromagnets on anisotropic triangular lattices. For
the isotropic triangular lattice model (TLM) the high-energy spectra
show several anomalous features that differ strongly from linear
spin-wave theory (LSWT). Even in the N\'{e}el phase, the deviations
from LSWT increase sharply with frustration, leading to roton-like
minima at special wavevectors. We argue that these results can be
interpreted naturally in a spinon language, and provide an
explanation for the previously observed anomalous finite-temperature
properties of the TLM. In the coupled-chains limit, quantum
renormalizations strongly enhance the one-dimensionality of the
spectra, in agreement with experiments on Cs$_2$CuCl$_4$.
\end{abstract}

\maketitle

One of the central problems in quantum magnetism is understanding
the properties of two-dimensional (2D) spin-$1/2$ Heisenberg
antiferromagnets (HAFM's). A question of particular interest is
whether the interplay between quantum fluctuations and geometrical
frustration can lead to unconventional ground states and/or
excitations. Candidate materials which have recently attracted much
attention include Cs$_2$CuCl$_4$ \cite{coldea} and
$\kappa$-(BEDT-TTF)$_2$Cu$_2$(CN)$_3$ \cite{shimizu}.

If the ground state is magnetically ordered, the system must have
gapless magnon excitations, which at sufficiently low energies are
expected to be well described by semiclassical (i.e., large-$S$)
approaches like spin-wave theory (SWT) and the nonlinear sigma model
(NLSM). However, if the magnon dispersion at higher energies
deviates significantly from the semiclassical predictions, it is
possible that the proper description of the excitations, valid at
all energies, is in terms of pairs of $S=1/2$ spinons. In this
unconventional scenario the magnon is a \textit{bound} state of two
spinons, lying below the two-spinon (particle-hole) continuum.

\begin{figure}[!htb]
\begin{center}
  \includegraphics[scale=0.45]{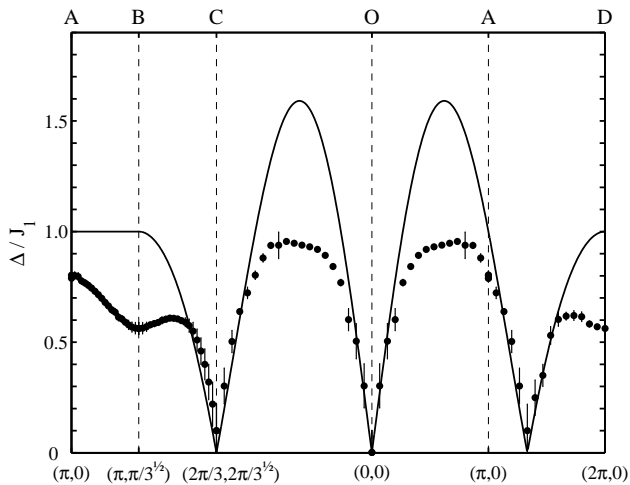}
  \caption{\label{fig_mk_y1}
Excitation spectrum for the TLM ($J_1=J_2$) along the path ABCOAQD
shown in Fig. \ref{trlattbz}. The high-energy spectrum is strongly
renormalized downwards compared to the LSWT prediction (full line).
Note the ``roton'' minima at B and D and the flat dispersion in the
middle parts of CO and OQ.}
\end{center}
\end{figure}

In this Letter we use series expansions to calculate the magnon
dispersion of 2D frustrated $S=1/2$ HAFM's. Our main finding is that
for the triangular lattice model (TLM) the dispersion shows major
deviations from linear SWT (LSWT) at high energies (Fig.
\ref{fig_mk_y1}). We argue that these deviations can be
qualitatively understood in terms of a two-spinon picture
\cite{anderson}, provided the spinon dispersion has minima at
$\bm{K}_i/2$, where $\bm{K}_i$ is a magnetic Bragg vector. Based on
this interpretation of the TLM spectra, we propose an explanation
for the anomalous finite-temperature behavior found in high
temperature series expansion studies \cite{elstner}.

Both qualitatively and quantitatively the deviations from LSWT found
here for the TLM are much more pronounced than those previously
reported \cite{singh95,sylju1,sandvik,zheng04} for the high-energy
spectra of the square lattice model (SLM). We point out that the
deviations from SWT increase in the N\'{e}el phase too, upon adding
frustration to the SLM. We further consider the limit of our model
relevant to Cs$_2$CuCl$_4$, and show that the calculated excitation
spectra are in good agreement with experiments \cite{coldea}.

\begin{figure}[!htb]
\begin{center}
\subfigure{\includegraphics[scale=0.38]{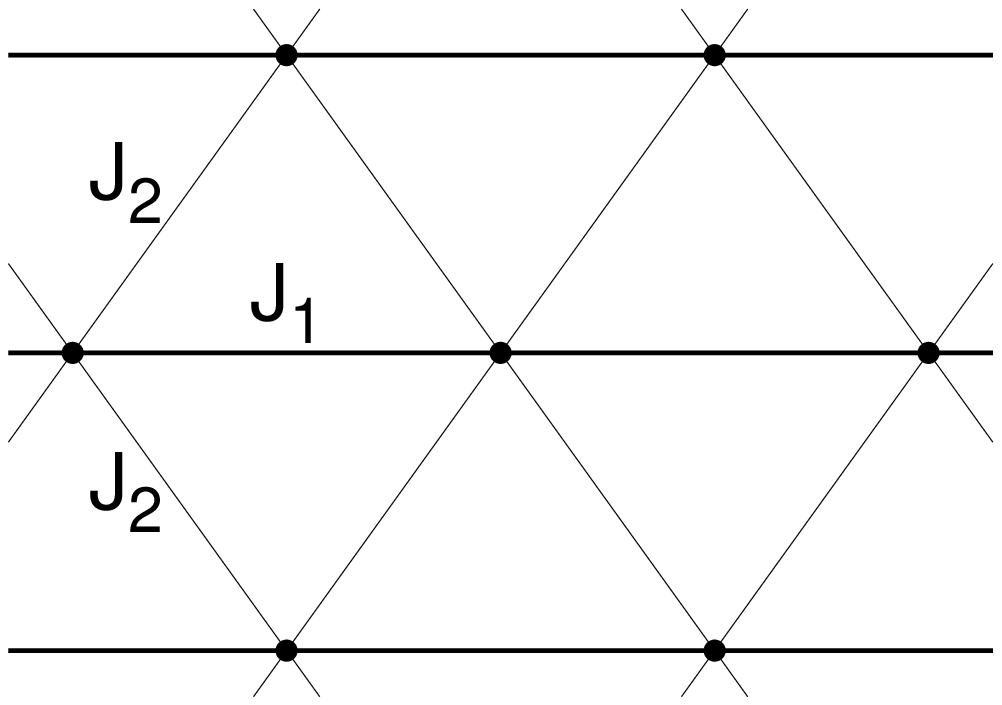}}
\subfigure{\includegraphics[scale=0.42]{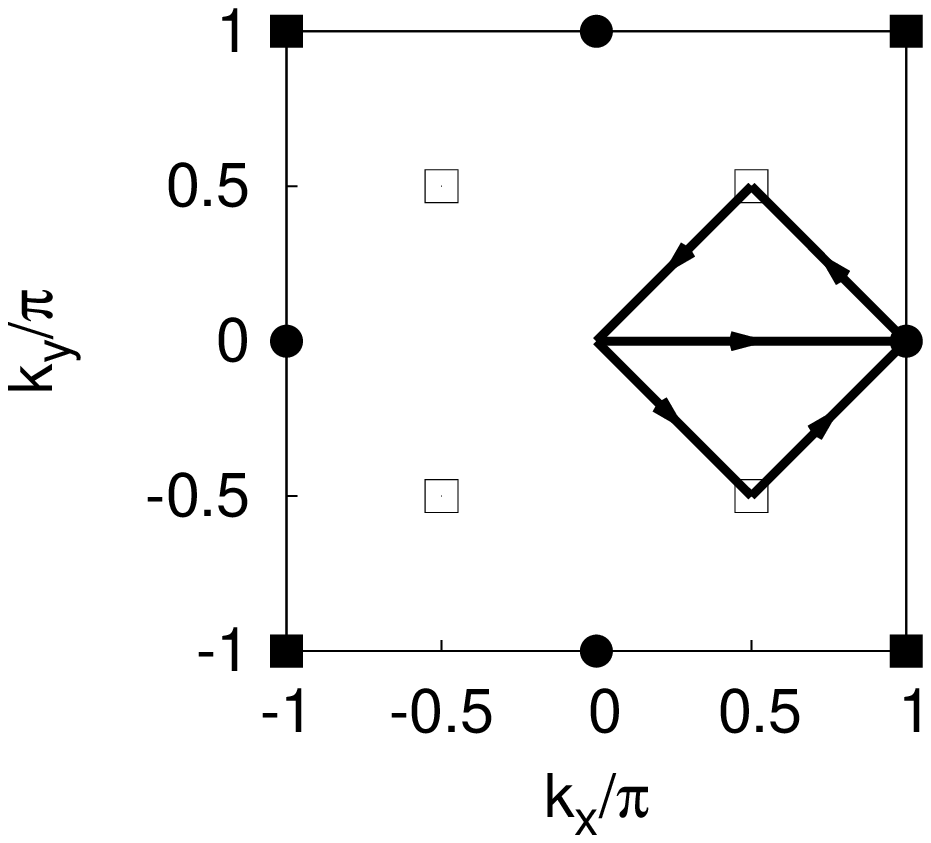}}
\caption{\label{Neelbz} (a) Exchange constants $J_1$ and $J_2$ for
the $S=1/2$ HAFM on the anisotropic triangular lattice. The model
can also be viewed as a square lattice with an extra exchange along
one diagonal. (b) Brillouin zone for the frustrated SLM, in standard
square lattice notation (the frustrating $J_1$ bonds are taken to
lie along the +45$^{\circ}$ directions in real space). The
excitation spectra in Fig. \ref{frustNeel} are plotted along the
bold path. Also shown are the locations of the Bragg vectors (filled
squares) and the local ``roton'' minima (filled circles) in the
$S=1$ dispersion, and the global minima of the $S=1/2$ spinon
dispersion (open squares) in the N\'{e}el phase. Note that the
90$^{\circ}$ rotation invariance present for $J_1=0$ is lost for
$J_1>0$.}
\end{center}
\end{figure}

\textit{Model.---}We consider a $S=1/2$ HAFM on an anisotropic
triangular lattice, with exchange couplings $J_1$ and $J_2$ (Fig.
\ref{Neelbz}(a)). This model interpolates between the SLM ($J_1=0$),
TLM ($J_1=J_2$), and decoupled chains ($J_2=0$). Classically the
model has N\'{e}el order for $J_1\le J_2/2$ with $q=\pi$, and
helical order for $J_1>J_2/2$ with $q=\arccos{(-J_2/2J_1)}$, where
$q$ ($2q$) is the angle between nearest-neigbor spins along $J_2$
($J_1$). The phase diagram for $S=1/2$ was studied in Refs.
\onlinecite{zheng99,chung}.

\textit{Series expansion method.---}In order to develop series
expansions for the helical phase (see Ref. \onlinecite{zheng99} for
expansions for the N\'{e}el phase spectra), we assume that the spins
order in the $xz$ plane, with nearest-neighbor angles $q$ and $2q$
as defined above; $q$ (generally different from the classical
result) is determined by minimizing the ground state energy. We now
rotate all the spins so as to make the ordered state ferromagnetic,
and introduce an anisotropy parameter in the Hamiltonian
$H(\lambda)=H_0 + \lambda V$, so that $H(0)$ is a ferromagnetic
Ising model and $H(1)$ is the spin-rotation invariant Heisenberg
model \cite{zheng99}. We use linked-cluster methods to develop
series expansions in powers of $\lambda$ for ground state properties
and the triplet excitation spectra. The calculation of the spectra
is particularly challenging as $S^z$ is not conserved. Due to
single-spin flip terms in $V$, the one-magnon state and the ground
state belong to the same sector, and the linked-cluster expansion
with the traditional similarity transformation \cite{gel96} fails.
To get a successful linked-cluster expansion, one has to use a
multi-block orthogonality transformation \cite{zhe01}. We have
computed the series for ground state properties to order
$\lambda^{11}$, and for the spectra to order $\lambda^{9}$ for
$J_1=J_2$ and to order $\lambda^8$ otherwise. The properties for
$\lambda=1$, discussed in the following, are obtained from standard
series extrapolation methods.

\textit{Square lattice model.---}Several numerical studies
\cite{singh95,sylju1,sandvik,zheng04} have reported deviations from
SWT for high-energy excitations in the SLM. While 1st order SWT
(i.e., LSWT) and 2nd order SWT predict no dispersion along
$(\pi-x,x)$, 3rd order SWT finds a weak dispersion, with the energy
at $(\pi,0)$ $\sim 2$\% lower than at $(\pi/2,\pi/2)$
\cite{zhengSWT,zheng04,igarashi05}. In contrast, the most recent
Quantum Monte Carlo (QMC) \cite{sandvik} and series expansion
\cite{zheng04} studies find the energy at $(\pi,0)$ to be $\sim 9$\%
lower. These deviations from SWT have been interpreted
\cite{hsu,sylju1,ho} in terms of a resonating valence-bond (RVB)
picture, in which the ground state is described as a $\pi$-flux
phase \cite{affmar} modified by correlations producing long-range
N\'{e}el order \cite{hsu}, and the Goldstone modes (magnons) are
bound states of a particle and a hole spinon \cite{hsu,ho}. The
magnon dispersion, calculated using a random phase approximation
(RPA), has local minima at $(\pi,0)$ \cite{hsu,ho} (in qualitative
agreement with the series/QMC results), the locations of which are
intimately related to the fact that the spinon dispersion has minima
at $(\pi/2,\pi/2)$ \cite{affmar,hsu}.

\begin{figure}[!htb]
\begin{center}
  \includegraphics[scale=0.42]{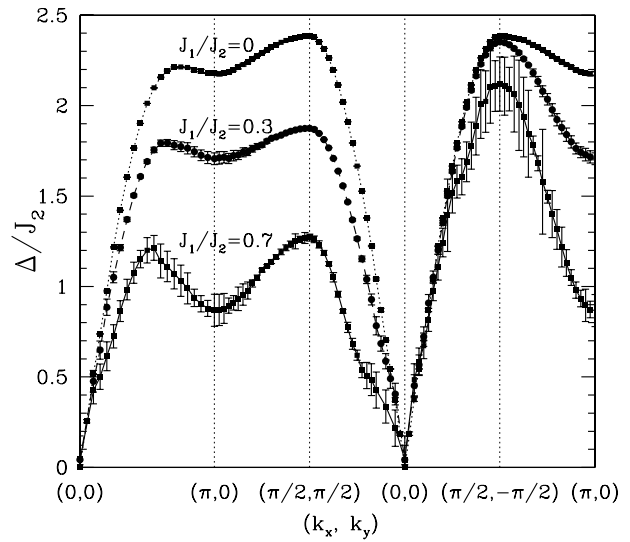}
  \caption{\label{frustNeel} Excitation spectra in the N\'{e}el
  phase. As $J_1/J_2$ is increased, the local ``roton'' minimum
  at $(\pi,0)$ becomes more pronounced, and the energy difference between
  $(\pi,0)$ and $(\pi/2,\pi/2)$, which LSWT predicts to be zero,
  increases.}
\end{center}
\end{figure}

\textit{Frustrated square lattice model.---}The N\'{e}el phase
persists up to $J_1/J_2=(J_1/J_2)_c\gtrsim 0.7$, after which the
system enters a dimerized phase \cite{zheng99}. As the frustration
$J_1/J_2$ is increased towards $(J_1/J_2)_c$, the local minimum at
$(\pi,0)$ becomes more pronounced (see Fig. \ref{frustNeel}); for
$J_1/J_2=0.7$ the energy difference between $(\pi/2,\pi/2)$ and
$(\pi,0)$ has increased to $\sim 31$\%. In contrast, LSWT predicts
no energy difference \cite{merino}. These results lend further
support to the RVB/flux-phase picture. The locations of the Bragg
vectors, roton minima, and spinon minima in the N\'{e}el phase are
shown in Fig. \ref{Neelbz}(b).

\begin{figure}[!htb]
\begin{center}
  \includegraphics[scale=0.6]{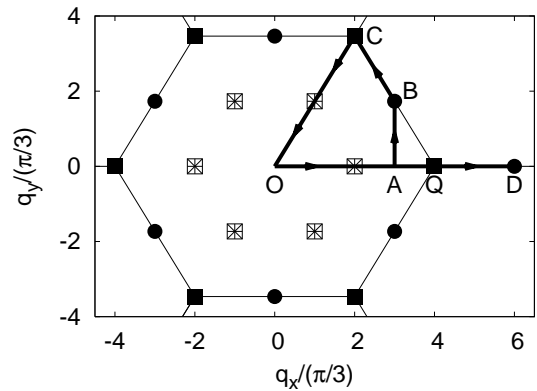}
   \caption{\label{trlattbz}
Brillouin zone diagram for the TLM. Bragg vectors (filled squares),
local roton minima (filled circles) and midpoints of flat dispersion
regions (stars) for the spin-$1$ dispersion are shown, as well as
the global minima of the proposed $S=1/2$ spinon dispersion (open
squares). Note that the latter two sets of wavevectors coincide. The
excitation spectra in Figs. \ref{fig_mk_y1} and \ref{fig_mk_y3} are
plotted along the path ABCOAD.}
\end{center}
\end{figure}

\textit{Triangular lattice model.---}Next, we consider the TLM
($J_1=J_2$), whose ground state has $120^{\circ}$ ordering between
neighboring spins \cite{order}. Fig. \ref{fig_mk_y1} shows the
excitation spectrum plotted along the path ABCOQD in Fig.
\ref{trlattbz}. While the low-energy spectrum near the (magnetic)
Bragg vectors looks conventional, the high-energy part of the
excitation spectrum shows several anomalous features which both
qualitatively and quantitatively differ strongly from LSWT: (i) At
high energies the spectrum is renormalized downwards with respect to
LSWT. (ii) The excitation spectrum is very flat in the region
halfway between the origin and a Bragg vector, with the midpoint
renormalized downwards with respect to LSWT by $\sim 40$\%. (iii)
There are local, roton-like, minima at the midpoints of the
Brillouin zone edges, whose energy is $\sim 44$\% lower than the
LSWT prediction. These strong downward renormalizations for the TLM
should be contrasted with the SLM, for which quantum fluctuations
always renormalize the LSWT spectrum \textit{upwards} and by amounts
never exceeding $20$\%.

\textit{Two-spinon interpretation.---}We will argue that these
results for the TLM are suggestive of spinons in the model. Our
basic hypothesis is that an ``uncorrelated'' RPA-like calculation
for the TLM, analogous to that discussed in Ref.
\onlinecite{schrieffer} for the SLM, should produce a spectrum
similar to the LSWT result, but it will be modified by correlations
\cite{seehsu}. In particular, repulsion from the two-spinon
(particle-hole) continuum can lower the magnon energy, especially at
wavevectors where this continuum has minima. These minima should
occur at $(\bm{K}_i-\bm{K}_j)/2$ corresponding to the creation of
minimum-energy particle-hole excitations with particle and hole
wavevectors $\bm{K}_i/2$ and $\bm{K}_j/2$, respectively, the
locations of which are shown in Fig. \ref{trlattbz} (the $\bm{K}_i$
are magnetic Bragg vectors). This equals (see Fig. \ref{trlattbz}) a
`roton' wavevector when $\bm{K}_i$ and $\bm{K}_j$ differ by a
$2\pi/3$ rotation around the origin, and a wavevector at a spinon
minimum if $\bm{K}_i$ and $\bm{K}_j$ differ by a $\pi/3$ rotation.
Fig. \ref{fig_mk_y1} shows that at both types of wavevectors the
excitation spectrum is strongly renormalized downwards with respect
to the LSWT. At the former (latter) type of wavevector, the LSWT
dispersion is flat (peaked), which upon renormalization leads to a
dip (flat region) in the true spectrum. Thus we attribute these
deviations to the existence of a two-spinon continuum.

For the SLM, the spectral weight of the magnon peak at $(\pi,0)$ is
considerably smaller than at $(\pi/2,\pi/2)$ (60\% vs 85\%)
\cite{sandvik}, and the magnon energy deviates much more from SWT at
$(\pi,0)$ than at $(\pi/2,\pi/2)$. This suggests quite generally
that the relative weight of the magnon peak decreases with
increasing deviation between the true magnon energy and the LSWT
prediction. Therefore one might expect the contribution of the
two-spinon continuum to be considerably larger for the TLM than for
the SLM.

As for the spinons proposed for the SLM \cite{affmar,hsu,ho}, the
locations of the minima in the spinon dispersion reflect a $d$-wave
character of the underlying RVB pairing correlations. A $d$-wave RVB
state of this type, but without long-range order, was discussed for
the TLM in Ref. \onlinecite{leefeng}. Its energy was however notably
higher than that of the ordered ground state, and attempts to modify
this RVB state to incorporate long-range order were not successful.
A mean-field RVB state for the TLM with \textit{bosonic} spinons,
whose dispersion has minima close to $\bm{K}_i/2$, was considered in
Refs. \onlinecite{yoshioka,lefmann}; again without long-range order.
In light of our spinon hypothesis for the TLM it would clearly be of
interest to revisit these problems.

\textit{Explanation of finite-temperature anomalies.---}The
existence of `roton' minima and their description in terms of pairs
of spinons provide a possible explanation for the sharply different
temperature dependent properties of the SLM and TLM. For the SLM the
temperature dependence of the correlation length is consistent with
a NLSM description in the renormalized classical (RC) regime over a
considerable temperature range \cite{chakravarty}. Even though the
ground state moment and spin-stiffness are comparable in the two
models, for the TLM the correlation length was found to be orders of
magnitude smaller at $T=J/4$ with a spin stiffness decreasing with
decreasing temperature (and longer length scales) \cite{elstner},
inconsistent with the NLSM description in the RC regime
\cite{nlsmtr}. We suggest that these differences  are due to the
fact that (see Figs. \ref{fig_mk_y1} and \ref{frustNeel}) the spinon
gap $E_s$ (which is half the roton energy) is four times smaller for
the TLM than for the SLM ($0.28 J$ versus $1.1 J$). Substantial
thermal excitation of spinons for temperatures comparable to $E_s$
will make a significant contribution to the entropy and reduce the
spin stiffness. Following an argument by Ng \cite{ng}, we expect
that thermal excitation of spinons will cause a NLSM description to
break down when $T \sim E_s$. The results of the high-temperature
expansions for both models are consistent with this estimate
\cite{elstner}.

\begin{figure}[!htb]
\begin{center}
  \includegraphics[scale=0.45]{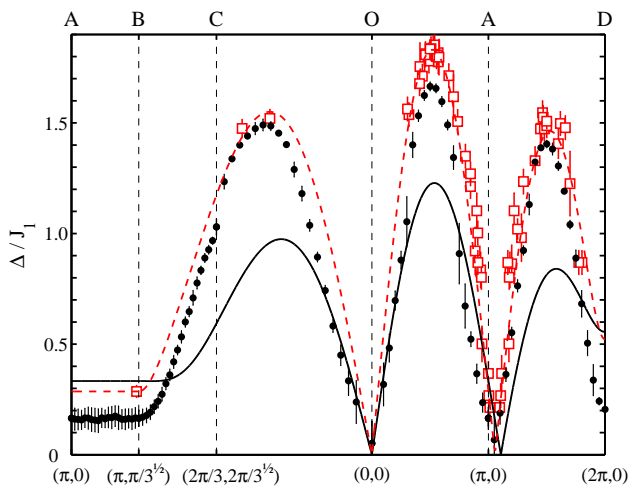}
  \caption{\label{fig_mk_y3}
Excitation spectrum for $J_1/J_2=3$ (solid points from series) along
the path ABCOAD in Fig. \ref{trlattbz}, compared to experimental
dispersion in Cs$_2$CuCl$_4$ (squares from Ref. \onlinecite{coldea},
dashed line is experimental fit) where the exchange ratio is
similar, $J_2/J_1=0.34(3)$ and $J_2=0.128(5)$ meV. Compared to LSWT
for $J_1/J_2=3$ (full line) these spectra are enhanced along the
$J_1$ bonds and decreased perpendicular to them.}
\end{center}
\end{figure}

\textit{$J_1/J_2=3$ and Cs$_2$CuCl$_4$.---}The ratio $J_1/J_2=3$,
closer to the decoupled chains limit, is relevant for
Cs$_2$CuCl$_4$, which has an extremely rich excitation spectrum
\cite{coldea} with well-defined spin-waves at low energies below the
N\'{e}el temperature $T_N$, and a continuum, strongly reminiscent of
the two-spinon continuum in one-dimensional (1D) antiferromagnets,
which persists well above $T_N$. Compared to LSWT the series
dispersion (Fig. \ref{fig_mk_y3}) is enhanced by $\sim53$\% along
the chains (close to the value for 1D chains) and reduced by $\sim
50$\% perpendicular to the chains. (In contrast, higher order SWT
\cite{veillette} gives only weak enhancement ($\sim 13$\%) over LSWT
along the chains.) Thus quantum fluctuations make the system appear
much more 1D. Overall, the series dispersion agrees well with the
experimental dispersion for Cs$_2$CuCl$_4$, also shown in Fig.
\ref{fig_mk_y3}, whose enhancement and reduction factors (derived
using a slightly different $J_1/J_2$ ratio; see caption) are
$\sim63$\% and $\sim17$\%. The large difference between the
theoretical and experimental reduction factors perpendicular to the
chains may be due to the Dzyaloshinski-Moriya interaction in
Cs$_2$CuCl$_4$ (not included in the series calculations) which may
make the system less 1D (it has the same path as $J_2$ and a
coupling constant $\sim0.15 J_2$) \cite{veillette}.

\textit{Acknowledgements.---}We thank J. P. Barjaktarevic, F. Becca,
S. Hayden, D. McMorrow, B. Powell, S. Sorella, and M. Veillette for
helpful discussions. This work was supported by the Australian
Research Council (WZ, JOF, and RHM), the US National Science
Foundation, Grant No. DMR-0240918 (RRPS), and the United Kingdom
Engineering and Physical Sciences Research Council, Grant No.
GR/R76714/01 (RC). We thank the Rudolf Peierls Centre for
Theoretical Physics at Oxford University (JOF), UC Davis, ISIS,
Rutherford Appleton Laboratory, and the Clarendon Laboratory at
Oxford University (RHM) for hospitality. We are grateful for the
computing resources provided by the Australian Partnership for
Advanced Computing (APAC) National Facility and by the Australian
Centre for Advanced Computing and Communications (AC3).


\begin{thebibliography}{99}

\bibitem{coldea} R. Coldea \textit{et al.}, Phys. Rev. B
\textbf{68}, 134424 (2003).

\bibitem{shimizu} Y. Shimizu \textit{et al.}, Phys. Rev. Lett.
\textbf{91}, 107001 (2003).

\bibitem{anderson} P. W. Anderson, Mater. Res. Bull. {\bf 8}, 153
(1973); P. Fazekas and P. W. Anderson, Philos. Mag. {\bf 30}, 423
(1974); P. W. Anderson, Science {\bf 235}, 1196 (1987).

\bibitem{elstner} N. Elstner \textit{et al.}, Phys. Rev.
Lett. \textbf{71}, 1629 (1993); J. Appl. Phys. \textbf{75}, 5943
(1994).

\bibitem{singh95} R. R. P. Singh and M. P. Gelfand, Phys. Rev. B
\textbf{52}, R15695 (1995).

\bibitem{sylju1} O. F. Sylju{\aa}sen and H. M. R{\o}nnow, J. Phys.
Condens. Matter \textbf{12}, L405 (2000).

\bibitem{sandvik} A. W. Sandvik and R. R. P. Singh, Phys. Rev. Lett.
\textbf{86}, 528 (2001).

\bibitem{zheng04} W. Zheng \textit{et al.}, Phys. Rev. B
\textbf{71}, 184440 (2005).

\bibitem{zheng99} Z. Weihong \textit{et al.}, Phys. Rev. B
\textbf{59}, 14367 (1999).

\bibitem{chung} C. H. Chung \textit{et al.}, J. Phys.: Condens.
Matter \textbf{13}, 5159 (2001).

\bibitem{gel96} M. P. Gelfand, Solid State Commun. \textbf{98},
11 (1996).

\bibitem{zhe01} W. Zheng \textit{et al.}, Phys. Rev. B
\textbf{63}, 144410 (2001).

\bibitem{zhengSWT} Z. Weihong and C. J. Hamer, Phys. Rev. B \textbf{47}, 7961 (1993).

\bibitem{igarashi05} J. Igarashi and T. Nagao, Phys. Rev. B \textbf{72}, 014403
(2005).

\bibitem{hsu} T. C. Hsu, Phys. Rev. B \textbf{41}, 11379 (1990).

\bibitem{ho} C. M. Ho \textit{et al.}, Phys. Rev. Lett.
\textbf{86}, 1626 (2001).

\bibitem{affmar} I. Affleck and J. B. Marston, Phys. Rev. B
\textbf{37}, R3774 (1988).

\bibitem{merino} J. Merino \textit{et al.}, J. Phys.: Condens.
Matter \textbf{11}, 2965 (1999).

\bibitem{order} B. Bernu \textit{et al.}, Phys. Rev. B \textbf{50},
10048 (1994); R. R. P. Singh and D. A. Huse, Phys. Rev. Lett.
\textbf{68}, 1766 (1992); D. J. J. Farnell \textit{et al.}, Phys.
Rev. B \textbf{63}, 220402(R) (2001); L. Capriotti \textit{et al.},
Phys. Rev. Lett. \textbf{82}, 3899 (1999).

\bibitem{schrieffer} J. R. Schrieffer \textit{et al.}, Phys. Rev.
B \textbf{39}, 11663 (1989).

\bibitem{seehsu} See also discussions of Eq. (2.44) and Fig. 2 in
Ref. \onlinecite{hsu}.

\bibitem{leefeng} T. K. Lee and S. Feng, Phys. Rev. B. \textbf{41},
11110 (1990).

\bibitem{yoshioka} D. Yoshioka and J. Miyazaki, J. Phys. Soc. Jpn.
\textbf{60}, 614 (1991).

\bibitem{lefmann} K. Lefmann and P. Hedeg{\aa}rd, Phys. Rev. B
\textbf{50}, 1074 (1994).

\bibitem{chakravarty} S. Chakravarty \textit{et al.}, Phys. Rev.
B \textbf{39}, 2344 (1989).

\bibitem{nlsmtr} P. Azaria \textit{et al.}, Phys. Rev. Lett.
\textbf{68}, 1762 (1992); A. V. Chubukov \textit{et al.}, Phys. Rev.
Lett. \textbf{72}, 2089 (1994).

\bibitem{ng} T. K. Ng, Phys. Rev. Lett. \textbf{82}, 3504 (1999).

\bibitem{veillette} M. Y. Veillette \textit{et al.}, Phys. Rev. B
\textbf{72}, 134429 (2005).

\end{thebibliography}
\end{document}